\documentclass[aps,twocolumn,showpacs,prr,amsmath,amssymb,floatfix,superscriptaddress]{revtex4-2}
\usepackage[colorlinks=true,citecolor=blue,linkcolor=RubineRed,urlcolor=blue]{hyperref} 
\usepackage[usenames,dvipsnames]{xcolor}
\usepackage{amsmath,amssymb,graphicx,color}
\usepackage{bm}
\usepackage{bbm}
\usepackage{float}
\usepackage[caption=false]{subfig}
\usepackage[usenames,dvipsnames]{xcolor}
\usepackage[normalem]{ulem}
\usepackage{soul}
\usepackage[normalem]{ulem}
\usepackage{soul,xcolor}
\usepackage{mathrsfs}

\usepackage{mathptmx} 
\DeclareMathAlphabet{\mathcal}{OMS}{cmsy}{m}{n}
\DeclareSymbolFont{largesymbols}{OMX}{cmex}{m}{n}

\begin{document}

\title{Topological Unwinding in an Exciton-Polariton Condensate Array}

\author{Guitao Lyu}\email{guitao@zju.edu.cn}
\affiliation{School of Physics and Zhejiang Institute of Modern Physics, Zhejiang University, Hangzhou, Zhejiang 310027, China}
\affiliation{Division of Natural and Applied Sciences, Duke Kunshan University, Kunshan, Jiangsu 215300, China}

\author{Yuki Minami}\email{yminami125@gmail.com}
\affiliation{Faculty of Engineering, Gifu University, 1-1 Yanagido, Gifu 501-1193, Japan}
\affiliation{School of Physics and Zhejiang Institute of Modern Physics, Zhejiang University, Hangzhou, Zhejiang 310027, China}

\author{Na Young Kim}\email{nayoung.kim@uwaterloo.ca}
\affiliation{Institute for Quantum Computing, University of Waterloo, 200 University Ave. West, Waterloo, Ontario, Canada, N2L 3G1}
\affiliation{Department of Electrical and Computer Engineering, University of Waterloo, 200 University Ave. West, Waterloo, Ontario, Canada, N2L 3G1}
\affiliation{Waterloo Institute for Nanotechnology, University of Waterloo, 200 University Ave. West, Waterloo, Ontario, Canada, N2L 3G1}

\author{Tim Byrnes}\email{tim.byrnes@nyu.edu}
\affiliation{New York University Shanghai; NYU-ECNU Institute of Physics at NYU Shanghai; Shanghai Frontiers Science Center of Artificial Intelligence and Deep Learning, 567 West Yangsi Road, Shanghai, 200126, China.}
\affiliation{State Key Laboratory of Precision Spectroscopy, School of Physical and Material Sciences, East China Normal University, Shanghai 200062, China}
\affiliation{Center for Quantum and Topological Systems (CQTS), NYUAD Research Institute, New York University Abu Dhabi, UAE.}
\affiliation{Department of Physics, New York University, New York, NY 10003, USA}

\author{Gentaro Watanabe}\email{gentaro@zju.edu.cn}
\affiliation{School of Physics and Zhejiang Institute of Modern Physics, Zhejiang University, Hangzhou, Zhejiang 310027, China}
\affiliation{Zhejiang Province Key Laboratory of Quantum Technology and Device, Zhejiang University, Hangzhou, Zhejiang 310027, China}

	\date{\today}

	\maketitle

\noindent{\large\textbf{Abstract}}\\ 
The phase distribution in a Bose-Einstein condensate can realize various topological states classified by distinct winding numbers.  While states with different winding numbers are topologically protected in the linear Schr{\"o}dinger equation, when nonlinearities are introduced, violations of the topological protection can occur, leading to unwinding. Exciton-polariton condensates constitute a nonlinear open-dissipative system that is well suited to studying such physics. Here we show that a one-dimensional array of exciton-polariton condensates displays a spontaneous phase unwinding from a $\pi$- to zero-state. We clarify that this collective mode transition is caused by the combined effect of nonlinearity and topological defects in the condensates. While the mode-switching phenomenon observed in our previous experiment was interpreted as the single-particle mode competition, we offer an alternative explanation in terms the collective phase unwinding and find its evidence by reanalyzing the experimental data. Our results open a route towards active control of the mode switching by manipulating the topological defects in prospective quantum polaritonic devices.

\bigskip
\noindent{\large\textbf{Introduction}}\\
The exciton-polariton (or polariton for short) is a quasiparticle existing in semiconductor microcavities, which consists of an exciton coupled with a cavity photon~\cite{Keeling2007review, Deng2010RMP, Carusotto2013RMP, Byrnes2014natphys, Deveaud2015review}. Because of their small effective mass, polaritons can form a Bose-Einstein condensate (BEC) at high temperatures, even at room temperature~\cite{Christopoulos2007PRL, Christmann2008APL, Baumberg2008PRL, Kena-Cohen2010natphot, Lu2012optexp, Li2013PRL, Plumhof2014natmat, Jiang2014PRX, Dietrich2016sciadv, Su2017nanolett, Lerario2017natphys, Zasedatelev2019natphot, Su2020natphys, Zasedatelev2021nature, Su2021natmat, Wei2022natcommun}. Although the lifetime of the polariton is very short, the condensate can be formed and maintained in a nonequilibrium manner when the loss is compensated by the laser pump. Such a driven-dissipative nature of the system makes the polariton condensate as a promising platform for the study of the nonequilibrium many-body phenomena~\cite{Bloch2022natrevphys}. Recently, active research on the Kardar-Parisi-Zhang dynamics~\cite{Fontaine2022nature, Gladilin2014PRA, Ji2015PRB, Altman2015PRX, He2015PRB, Wachtel2016PRB, He2017PRL, Zamora2017PRX, Squizzato2018PRB, Diessel2022PRL} and the Kibble-Zurek mechanism~\cite{Matuszewski2014PRB, Solnyshkov2016PRL, Kulczykowski2017PRB, Zamora2020PRL, Solnyshkov2021PRB} has been going on with this platform.

Periodic external potentials for the exciton-polariton system have been realized by various methods~\cite{Schneider2017RepProgPhys}, such as the deposition of metal strips on the surface of the microcavity~\cite{Lai2007nature, Utsunomiya2008natphys, Kim2008pssb}, etching of micropillars~\cite{Gutbrod1998PRB, Bayer1999PRL, Tanese2013natcommun, Jacqmin2014PRL}, and others~\cite{Lima2006PRL, Cerda-Mendez2010PRL, Winkler2015NJP, Zhang2015PNAS}. This development has uncovered avenues for applications of exciton-polariton systems to the simulation of driven-dissipative quantum many-body systems; similarly to the optical lattice for cold atomic gases, which allowed a unique ability to mimic the solid-state physics models opening a research field of quantum simulations~\cite{Lewenstein2012book, Gross2017science}. One interesting direction that has been pursued in particular is to construct various types of topological quantum states in controllable quantum systems \cite{ozawa2019topological,zhang2018topological}.  Examples and applications include one-dimensional (1D) optical cavity arrays \cite{luo2015quantum}, topological quantum emitters \cite{barik2018topological,ishida2022large}, quantum simulation of topological states in photonic waveguide arrays \cite{ke2016topological,lustig2019photonic}, the momentum-space lattices of atomic BECs \cite{meier2016observation,xie2019npj}, and demonstrations of topological quantum computing \cite{huang2021emulating}.

In ref.~\cite{kanamoto2008topological}, Kanamoto, Carr, and Ueda found a nontrivial mechanism for winding and unwinding of topological quantum states in a 1D toroidal BEC. Despite the common perception that there is a discontinuous jump in energy between states of different winding number in such a system, it was found that the nonlinearity of BECs allows for a continuous transition between excited metastable states, originating from soliton trains bifurcating from the plane-wave solution. Regarding the exciton-polariton system, on the other hand, given its driven-dissipative nature in contrast to the system considered in ref.~\cite{kanamoto2008topological} with particle-number conservation, it is an intriguing question of whether and how metastable relaxation in the nonlinear collective dynamics can account for mode transitions.
In ref.~\cite{Lai2007nature}, Lai, Yamamoto, and co-workers including two of us observed the coherent $\pi$-state (a state at the edge of the first Brillouin zone with the phase difference by $\pi$ between neighbouring sites) and zero-state (ground state) of the polariton condensate in a weak periodic potential. Further, they discovered a mode-switching phenomenon between the $\pi$- and zero-states identified in a pump-power dependence of the emission intensity of photons from these states, which was interpreted as the \textit{mode competition} in their discussion~\cite{Lai2007nature}. This experimental result sparked active research on the dynamics of polariton condensates in a periodic potential~\cite{Krizhanovskii2009PRB, Krizhanovskii2013PRB, Ma2015PRB, Chestnov2016PRB, Charukhchyan2016JPCS, Winkler2016PRB, Yoon2019PRA, Moilanen2022PRB}.
Up to this point, the mode-switching phenomenon has been interpreted as mode competition based on single-particle phenomenon~\cite{Lai2007nature, Deng2010RMP}. However, the actual microscopic process behind the mode-switching phenomenon is yet to be clarified whether it is collective condensate dynamics~\cite{Winkler2016PRB} or single-particle transition~\cite{Lai2007nature, Deng2010RMP}.
Understanding this mechanism is an important step toward the active control of the polariton condensates in a periodic potential.

In this paper, we present a comprehensive scenario for the mode-switching phenomenon observed in ref.~\cite{Lai2007nature} based on the collective mode transition. By incorporating the collective dynamics of the condensate wave function and the single-particle transition on equal footing, we reveal that the mode-switching phenomenon occurs through topological phase unwindings of the polariton condensate wave function, which results from the combined effects of nonlinearity and topological defects of the condensate.
This is in contrast to the existing scenario of the mode switching based on the single-particle processes conjectured in the original experimental work~\cite{Lai2007nature, Deng2010RMP}. By reanalyzing the experimental data of ref.~\cite{Lai2007nature}, we find an evidence of the collective phase unwinding.

\bigskip
\noindent{\large\textbf{Results and Discussion}}

\bigskip
\noindent\textbf{Model.}
We consider polariton condensates in a 1D periodic potential $V(x)=V_0 \cos(2\pi x/d)$ in the $x$ direction, where $V_0$ is the potential depth and $d$ is the lattice constant. This system can be described by the generalized Gross-Pitaevskii equation (gGPE)~\cite{Keeling2008PRL, Eastham2008PRB, Wouters2010PRL, Keeling2011Contempphys, Moxley2016PRA}:
\begin{equation}
	\begin{aligned}
		i \hbar\frac{\partial\Psi(x,t)}{\partial t} =\Big[  &-\dfrac{\hbar^2}{2m} \dfrac{\partial^2}{\partial x^2} + V_0\cos \Big( \frac{2\pi}{d}x \Big) +g|\Psi(x,t)|^2 \\
		&+\frac{i}{2}\left(P - \gamma - \eta|\Psi(x,t)|^2\right) \Big] \Psi(x,t).
	\end{aligned}
	\label{eq:ggpe}
\end{equation}
Here $\Psi$ is the condensate wave function of polaritons, $m$ is the effective mass of the polariton, $g$ is the two-body interaction strength of the polaritons in the condensates, $P$ is the gain rate of the polariton condensate which is determined by the laser pump power, and $\gamma$ is the loss rate due to the leakage of photons in the microcavity. The gain saturation effect due to the existence of reservoir excitons is phenomenologically described by the $\eta$ term \cite{Keeling2008PRL}.

For all the calculations presented in the paper, we take $E_{0}=\hbar^{2}\pi^{2}/(2md^{2})$ as the unit of energy and $\tau=\hbar/E_{0}$ as the unit of time. As in the experiment~\cite{Lai2007nature}, we set $m \approx 9\times10^{-5}m_\text{e}$, where $m_\text{e}$ is the electron mass, and $d = 2.8\ \mu\text{m}$; thus, $E_{0} \approx 0.533\ \text{meV}$ and consequently $\tau \approx 1.24\ \text{ps}$. For other parameters, we set $g = 6 \times 10^{-3}\ \text{meV}\ \mu$m$^2$, $\gamma = 0.33\ \text{ps}^{-1}$, and $\eta = 0.01\ \text{ps}^{-1}\ \mu\text{m}^2$, which are consistent with the experiment~\cite{Lai2007nature} (see the parameter estimate in Methods).
Since the system size of the experiment~\cite{Lai2007nature} is the order of 10 (typically $20$--$30$) unit cells, the simulations presented in the paper employ the system with $10$ unit cells from $x=-5d$ to $5d$ under periodic boundary conditions to simulate the bulk region of the sample unless specified otherwise. We have also performed simulations for larger system size with a few hundreds unit cells, and have confirmed that the main results discussed in the paper do not change.

We have confirmed that another standard formalism, coupled driven-dissipative Gross-Pitaevskii (GP) equations~\cite{Wouters2007PRL, Carusotto2013RMP, Ma2015PRB, Chestnov2016PRB} which includes the evolution of the exciton reservoir, also gives qualitatively the same results of the mode-switching phenomenon. For the presentation in this paper, however, we opt to use the gGPE (\ref{eq:ggpe}) for clarity of the discussion since this simpler model already captures the essential physics of our problem.

\begin{figure}[tbp!]
	\centering
	\includegraphics[width=1\linewidth]{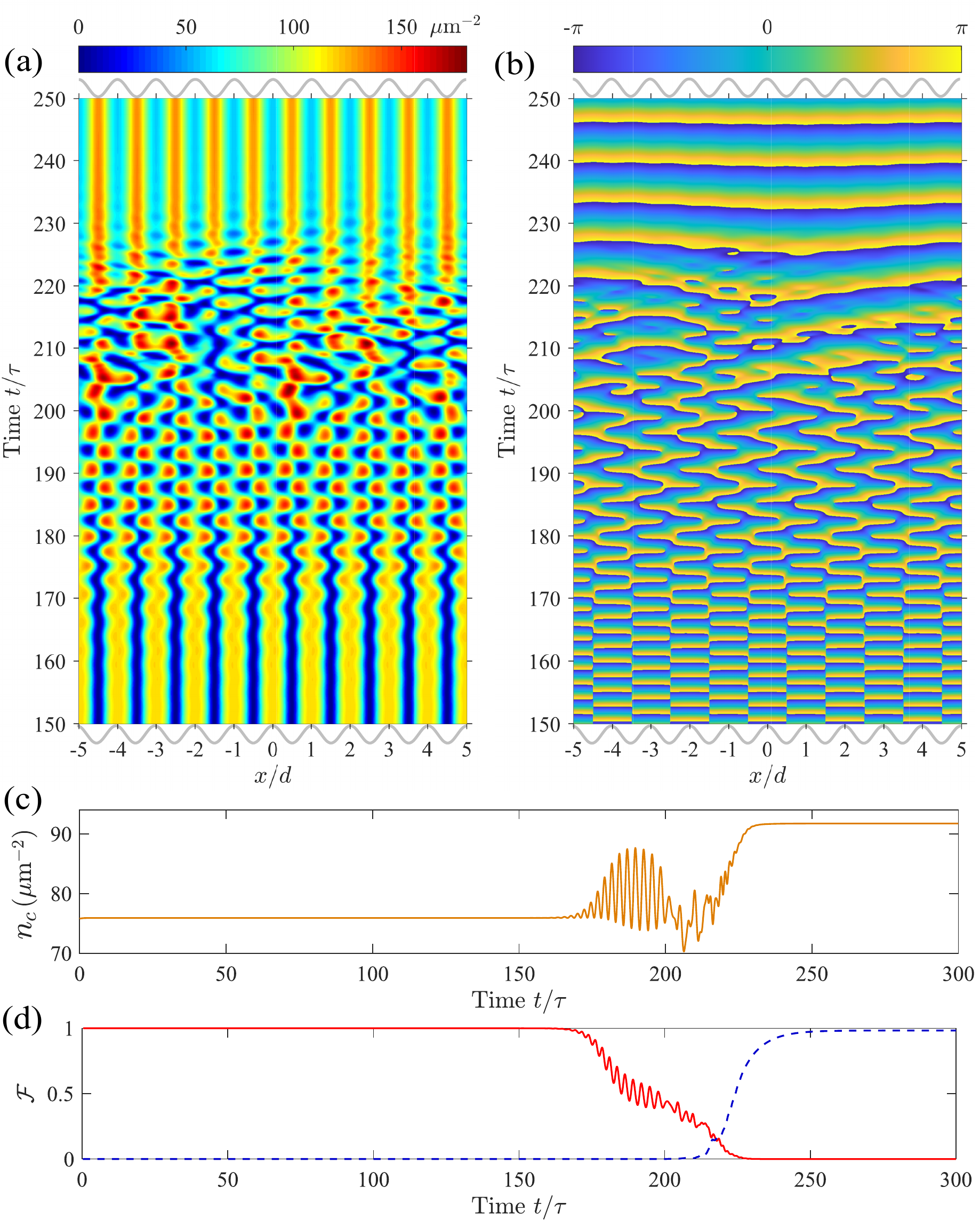}
	\caption{{\bf Dynamics of the transition from the $\pi$-state to the zero-state.} The time evolution starts from the $\pi$-state. (a) and (b) show the density $|\Psi(x,t)|^2$ and the phase $\Phi(x,t)$ of the condensate, respectively. The cosine curves at the bottom and top of these panels show the shape of the external potential $V(x)$. (c) shows the average condensate density $n_{\rm c}$. (d) presents the fidelity $\mathcal{F}_{i}$ between the normalized wave function $\psi(x,t)$ and the normalized $\pi$-state (red solid curve) or zero-state (blue dashed curve). Here we take $V_0/\hbar = 1\ \text{ps}^{-1}$ and $P/\gamma=4$.}
	\label{fig:modetrans}
\end{figure}

\bigskip
\noindent\textbf{Mode transition under constant pump.}
First, we study the dynamical process of the transition from the $\pi$-state to the
zero-state (referred to as ``mode transition'' hereafter) under an idealized situation with constant pump power $P$. For various values of $P$ and $V_0$, we have performed numerical simulations of the time evolution based on the gGPE (\ref{eq:ggpe}) starting from the $\pi$-state as the initial condition, which is obtained by solving the time-independent version of the gGPE (see Methods for details of the Bloch band structure). For sufficiently large values of $P$ such that the average density of the initial steady $\pi$-state is large enough, we have observed that the system undergoes a transition from the $\pi$-state to the
zero-state. The time evolution of the density profile $|\Psi(x,t)|^2$ and the phase profile $\Phi(x,t) \equiv \arg{[\Psi(x,t)]}$ of the condensate are displayed in Figs.~\ref{fig:modetrans}(a) and \ref{fig:modetrans}(b), respectively. Regarding the initial $\pi$-state, the density maxima are located at the maxima of the cosine potential $V(x)$, and the phase profile is a stepwise function consisting of a constant plateau region and a $\pi$-phase jump in each unit cell. In the beginning stage of the time evolution ($0 \le t\lesssim 160 \tau$), the system is almost static, and the average condensate density $n_{\rm c}$ is almost constant in time [Fig.~\ref{fig:modetrans}(c)]. From $t \approx 160 \tau$, the density profile starts to oscillate in the positive and negative $x$ directions [Fig.~\ref{fig:modetrans}(a)]. This is caused by the spontaneous growth of the perturbation from the initial $\pi$-state originated from small numerical errors. Such a spontaneous growth of the perturbation is a typical signature of the dynamical instability~\cite{Chestnov2016PRB, Baboux2018Optica}, which is a consequence of the nonlinear effect in the initial $\pi$-state due to its sufficiently large average density. After the period of the gradual increase of the oscillation amplitude, the system undergoes an almost regular large-amplitude oscillation for a while ($t \approx 180\, \text{--}\, 200 \tau$). Then, at $t \approx 200\, \text{--}\, 225 \tau$, the condensate density $|\Psi(x,t)|^2$ oscillates irregularly [Fig.~\ref{fig:modetrans}(a)] and phase-unwinding processes occur, through which the phase winding number of the condensate finally becomes zero as we discuss further later. After the unwinding of the phase, the system decays into the zero-state, whose phase is constant in space [see at $t \gtrsim 230 \tau$ in Fig.~\ref{fig:modetrans}(b)] and the density maxima are located at the potential minima.

To demonstrate that this case is indeed a mode transition from the $\pi$- to zero-state, as shown in Fig.~\ref{fig:modetrans}(d), we calculate the fidelity $\mathcal{F}_i$ between the normalized condensate wave function $\psi(x,t) \equiv \Psi(x,t)/\sqrt{n_{\rm c}(t)}$ and the normalized $\pi$-state $\varphi_{\pi}$ (red solid curve) or zero-state $\varphi_{0}$ (blue dashed curve), given by $\mathcal{F}_{i}= \left|\int dx\, \varphi^{*}_{i}(x)\, \psi(x,t) \right|^2$ with $i=\pi$ and $0$, respectively. Here, $\varphi_i$'s are normalized to unity: $\int dx\, |\varphi_i(x)|^2 = 1$. The mode transition observed here is a collective process through the condensate wave function, which is completely different from the one by the single-particle transition considered in ref.~\cite{Lai2007nature}.
\begin{figure}[tb!]  
	\centering\includegraphics[width=1\linewidth]{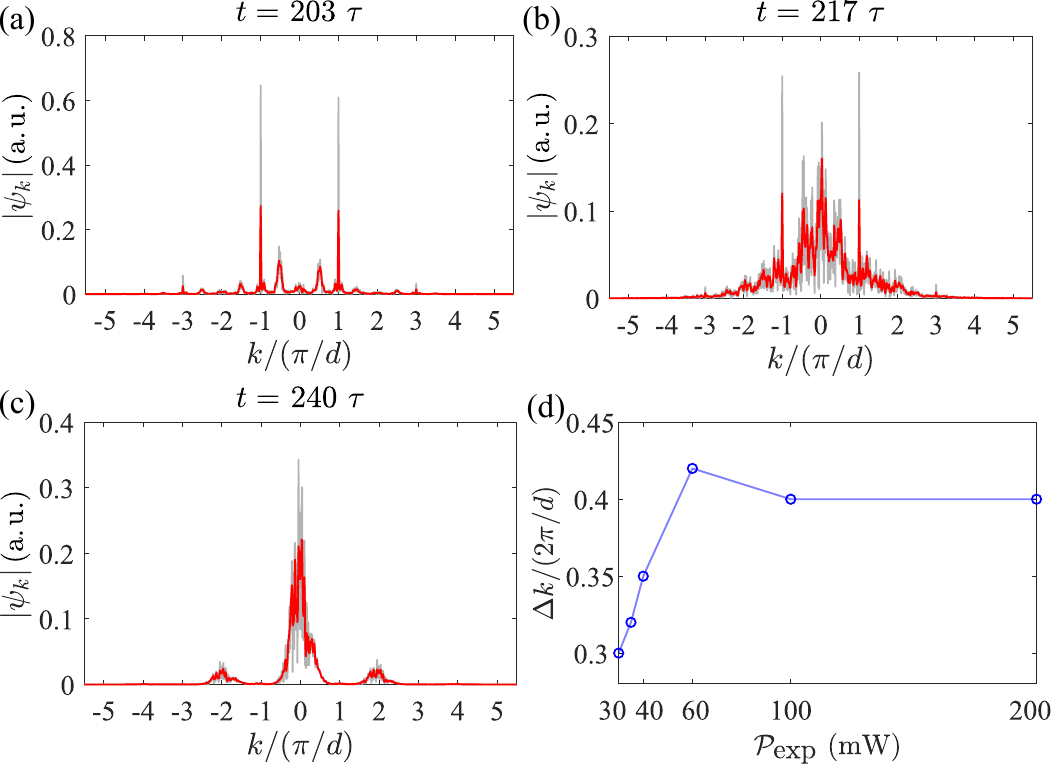}
	\caption{{\bf Broadened momentum distribution in the mode transition process and its experimental evidence.} (a)\,--\,(c) present the snapshots of the square root of the momentum distribution in the mode transition process. The light gray curves represent the raw data obtained by performing a fast Fourier transform of $\psi(x,t)$, and the dark red curves depict outcomes after the Gaussian smoothing with a full-width-at-half-maximum (FWHM) of $0.02\pi/d$, a resolution achievable in current experiments~\cite{Lai2007nature,Claude2022PRL}. (d) shows the FWHM $\Delta k$ of the zero-state peak around $k=0$ in Fig.~4(b) of ref.~\cite{Lai2007nature} for various pump power.}
	\label{fig:momdistr}
\end{figure}  

The momentum distribution of polaritons, which can be measured experimentally by angle-resolved photoluminescence spectroscopy~\cite{Houdre1994PRL,Lai2007nature,Utsunomiya2008natphys,Xie2012PRL,Claude2022PRL}, provides further information about the mode transition dynamics. We calculate the Fourier transform $\psi_k(t)$ of the instantaneous normalized condensate wave function $\psi(x,t)$ at different moments in time. For this calculation, we perform another simulation with a larger system size of $200$ unit cells to get better resolution in momentum space.
Figures~\ref{fig:momdistr}(a)\,--\,\ref{fig:momdistr}(c) show several representative snapshots of the square root of the momentum distribution, $|\psi_k(t)|$, during the time evolution.  The momentum distribution $|\psi_k(t)|^2$ does not change until $t \lesssim 160 \tau$ since the system is stationary in the first stage of the time evolution as mentioned previously. There are two main peaks located at $k=\pm \pi/d$, small peaks at $k=\pm 3 \pi/d$ and further smaller peaks at higher odd multiples of $\pi/d$ which are characteristic of the $\pi$-state. During the phase-unwinding stage ($t \approx 200-225 \tau$), many other $k$ components different from odd multiples of $\pi/d$ appear. When the system undergoes an irregular oscillation in the beginning of this stage, $k$ components with integer and half-integer multiples of $\pi/d$ become significant as seen in Fig.~\ref{fig:momdistr}(a). When many phase-unwinding processes occur in the middle of this stage, various other $k$ components including those with non-integer $k/(\pi/d)$ also appear and the momentum distribution becomes broad as presented in Fig.~\ref{fig:momdistr}(b). Finally, all $k$ components vanish except for those corresponding to the zero-mode, and results in a momentum distribution with a main peak at $k=0$ and smaller peaks at $k=\pm 2 \pi/d$ [Fig.~\ref{fig:momdistr}(c)].
The nonzero width of the peaks at $k=0$ and $\pm2 \pi/d$ in Fig.~\ref{fig:momdistr}(c) is due to the small density fluctuation around the zero state in the time evolution.

The emergence of the broad momentum distribution in the phase-unwinding stage [Fig.~\ref{fig:momdistr}(b)] is a striking signature of the mode transition by the collective dynamics of the condensate wave function. Such broad momentum distribution with various $k$ components including non-integer multiples of $\pi/d$ cannot be observed for the single-particle transition from the $\pi$-state to the zero-state considered in ref.~\cite{Lai2007nature}. In a single-particle transition scenario, there should be peaks only at $k=0$ and $\pm 2\pi/d$ corresponding to the zero-state, and odd multiples of $\pi/d$ corresponding to the $\pi$-state.

In actual experiments, measured momentum distribution is also broadened by two additional mechanisms. First, due to the finite size of the system, the momentum of the steady state has a width of $\sim 2\pi/L$ where $L$ is the size of the system. However, this width is much smaller than the spread $\sim 2\pi/d$ of the momentum distribution shown in Fig.~\ref{fig:momdistr}(b). Second, if the momentum distribution is measured by the diffraction pattern of the lower-polariton emission as in the experiment of ref.~\cite{Lai2007nature}, the peak of each mode in the momentum space spreads due to the photon diffraction. For example, the full-width-at-half-maximum (FWHM) of the diffraction peaks from the $\pi$-state and zero-state is about $0.3$ $\mu$m$^{-1}$  and $1.0$ $\mu$m$^{-1}$, respectively, in the case of ref.~\cite{Lai2007nature}. However, they are still smaller than the spread (FWHM) observed in Fig.~\ref{fig:momdistr}(b): $\sim 2\pi/d \approx 2.2$ $\mu$m$^{-1}$. Therefore, although the above mechanisms affect the measurement of the momentum distribution, the predicted broadening of the $k=0$ peak by the phase-unwinding is still observable.

Figure~\ref{fig:momdistr}(d) shows experimental measurements of the FWHM $\Delta k$ of the zero-state peak in the momentum distribution~\cite{Lai2007nature}. In this experiment, power ${\mathcal P}_{\rm exp}$ of the pump laser at the condensation threshold is $20$~mW. Above the threshold, the $\pi$-state is dominant until ${\mathcal P}_{\rm exp} \approx 40$~mW, and the zero-state starts to take over when ${\mathcal P}_{\rm exp} \gtrsim 40$~mW. In the collective transition scenario, $\Delta k$ is thus predicted to increase at ${\mathcal P}_{\rm exp} \approx 40$~mW. The behavior of $\Delta k$ shown in Fig.~\ref{fig:momdistr}(d) is consistent with this prediction, and this is strong evidence for the collective phase unwinding process behind the mode-switching phenomenon in the experiment~\cite{Lai2007nature}.

\begin{figure}[tb!]  
	\centering\includegraphics[width=1\linewidth]{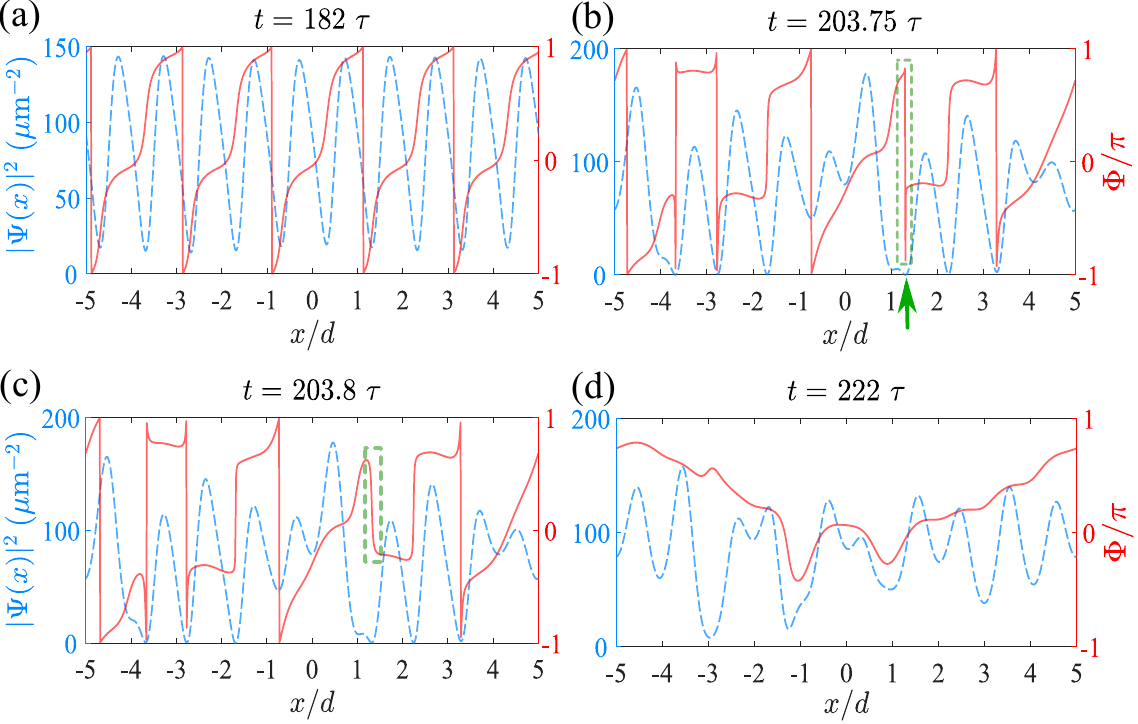}
	\caption{{\bf Phase-unwinding process in the mode transition.} (a)\,--\,(d) present the snapshots of the density profile (cyan dashed curve) and the phase profile (red solid curve) in the time evolution of Fig.~\ref{fig:modetrans}. The arrow in (b) marks the position with zero-density. The green dashed box in (b) and (c) highlights the phase unwinding.
	}
	\label{fig:unwind}
\end{figure}

\bigskip
\noindent\textbf{Transition mechanism.}
In the Schr\"odinger equation with periodic potential, the transition from the $\pi$- to zero-state is prohibited because of the quasimomentum conservation due to the periodicity of the system. In addition, the phase winding number of the state before the transition is topologically protected because the magnitude of the condensate wave function of this state is nonzero everywhere in the system with the periodic boundary condition [see, e.g., Fig.~\ref{fig:unwind}(a)]. However, in polariton condensates, the protection of the initial $\pi$-state by quasimomentum conservation is relaxed due to the nonlinearity of the system. If the initial $\pi$-state is dynamically unstable when the average density is sufficiently large, a growing perturbation can make the density profile aperiodic while keeping the initial quasimomentum (i.e., spatial phase variation rate) unchanged. However, once the density profile becomes aperiodic, the quasimomentum is no longer a conserved quantity since the nonlinear term $g|\Psi(x)|^2$ in gGPE (\ref{eq:ggpe}) plays a role as a part of the effective external potential. Furthermore, the topological protection of the phase winding can also be violated by the emergence of zeros of $\Psi(x)$ (i.e., topological defects). Since the phase of $\Psi$ is not well-defined and can take any value at the position of $\Psi(x)=0$, a discontinuous jump in the phase is allowed there. Consequently, the phase winding can be removed by the defect as we will see below.

Figure~\ref{fig:unwind} shows several representative snapshots of the phase and density profiles in the time evolution of Fig.~\ref{fig:modetrans}, clarifying the phase-unwinding process. Figure~\ref{fig:unwind}(a) is a snapshot at $t=182 \tau$ in the regular large-amplitude oscillation of the density. We can clearly observe that the phase (red solid curve) changes by $2\pi$ in each two unit cells and the density (cyan dashed curve) is still periodic and nonzero at any position. In the phase-unwinding stage which is triggered by the irregular large-amplitude oscillation, the density becomes aperiodic and zeros of $\Psi(x)$ are created frequently. Figure~\ref{fig:unwind}(b) captures a snapshot of the onset of the phase unwinding by a topological defect. Here, a topological defect at $x\approx 1.3 d$ marked by the green arrow just starts to remove the phase winding by $2\pi$. As highlighted by the green dashed box in Figs.~\ref{fig:unwind}(b) and \ref{fig:unwind}(c), when the phase winding is removed by the topological defect, the phase shows a discontinuous jump there and the phase is finally unwound by a multiple of $2\pi$. (Although the actual system in the experiment is not perfectly periodic but finite, the phase unwinding by topological defects observed in our simulation can still happen in the bulk region of the sample in the experiment.) After many such phase-unwinding processes, the winding number of the phase finally becomes zero as shown in Fig.~\ref{fig:unwind}(d). In addition, due to the dissipation, the condensate finally decays to the zero-state.

\bigskip
\noindent\textbf{Mode switching.}
Finally, we demonstrate that the experimental results of the mode-switching phenomenon~\cite{Lai2007nature} can be explained by the collective mode transition. Here, we perform simulations with pulsed pump as in the experiment. In the simulations, we also incorporate the single-particle transition process from the $\pi$- to zero-state in addition to the collective mode transition process, and we shall demonstrate that the collective transition is the major process predominant over the single-particle one.

First, we simulate the population dynamics of the $\pi$-state and the zero-state after the onset of a pump pulse, which has been measured in the experiment~\cite{Lai2007nature}. In the simulations, we first prepare the stationary $\pi$-state with a very small condensate density as an initial seed by setting the pumping strength just above the threshold ($P/\gamma = 1.15$). Then, we calculate the time evolution with a Gaussian pump pulse (see Methods) whose width and time-averaged strength $\overline{P}$ are consistent with the pump pulse used in the experiment~\cite{Lai2007nature}. To incorporate the contribution of the single-particle transition from the $\pi$- to zero-state, an extra term, which reduces to the term corresponding to this process in the rate equation employed in ref.~\cite{Lai2007nature}, is added to gGPE (\ref{eq:ggpe}) (see the last section in Methods for details). For the transition rate of this process, we use the same value as ref.~\cite{Lai2007nature}. Hereafter, we set $V_0/\hbar=0.1\ \text{ps}^{-1}$ which is comparable to the experimental value.

\begin{figure}[tpb!]
	\centering
	\includegraphics[width=1\linewidth]{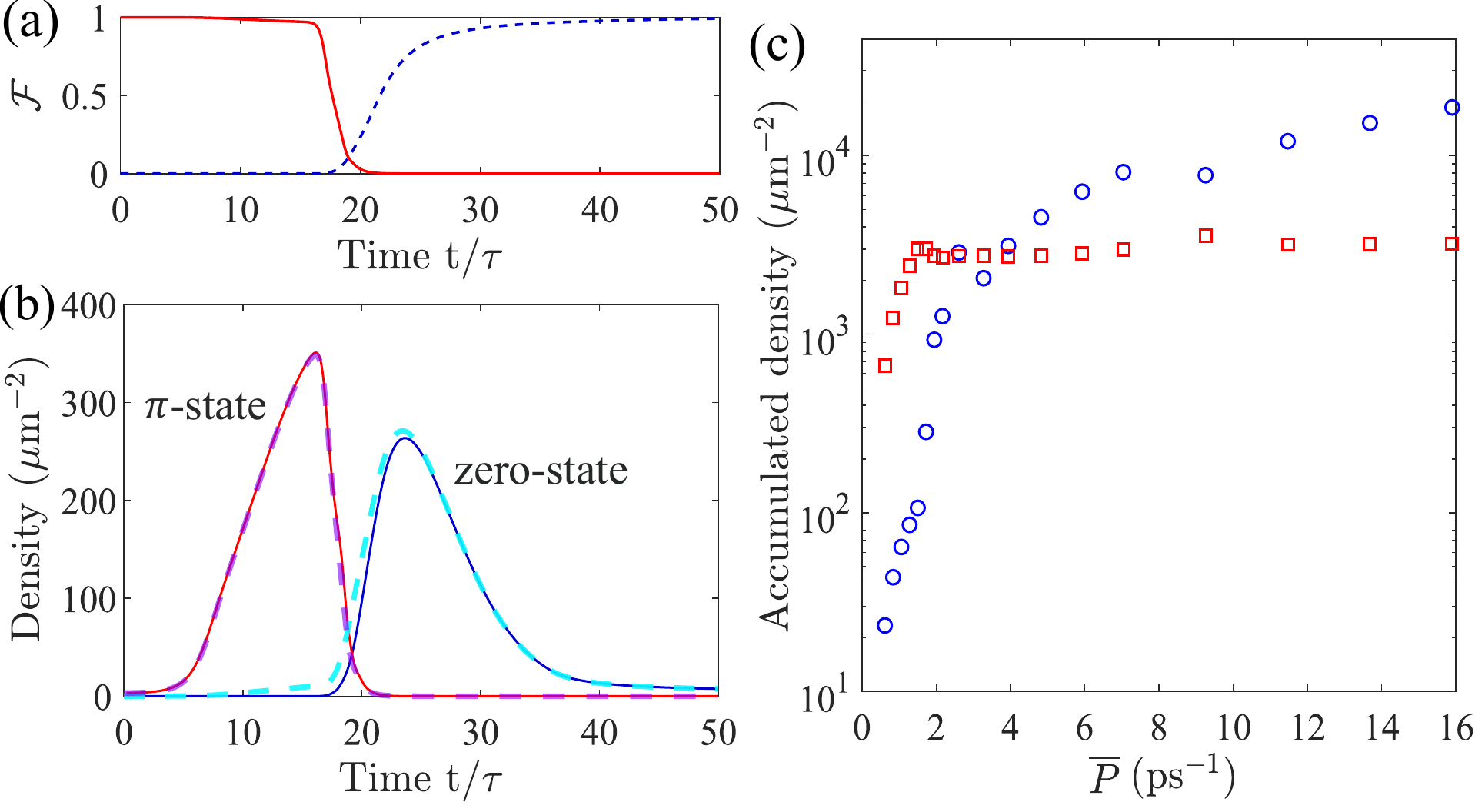}
	\caption{{\bf Simulation for the experimental results.}
		(a)\,--\,(b) Population dynamics of the $\pi$- and zero-states under the Gaussian pump pulse. (a) presents the fidelity $\mathcal{F}_{i}$ between the normalized $\Psi(x,t)$ and $\pi$- (red solid curve)  or zero-state (blue dashed curve) . (b) shows the time evolution of the average density of the $\pi$-state (first peak) and zero-state (second peak). The result with (without) the single-particle transition is shown by the dashed (solid) curves. (c) Mode-switching phenomenon. The accumulated density of the $\pi$- (red squares) and the zero-states (blue circles) is plotted as a function of the time-averaged pumping strength $\overline{P}$.
		Here we set $V_0/\hbar=0.1\ \text{ps}^{-1}$, and the averaged pumping strength for (a) and (b) is $\overline{P}/\gamma \approx7.9$.
	}
	\label{fig:modetrans_pulse}
\end{figure}

Figure \ref{fig:modetrans_pulse}(a) shows the fidelity between the normalized condensate wave function $\psi(x,t)$ and the normalized $\pi$-state $\varphi_{\pi}$ (red solid curve) or zero-state $\varphi_{0}$ (blue dashed curve) for the case without the single-particle transition. One can see that the mode transition happens in $t \approx 15-25\tau$ in this case. The population dynamics of the $\pi$- and zero-states is shown in Fig.~\ref{fig:modetrans_pulse}(b). Here, we plot the average density of the $\pi$-state and zero-state given by $n_{i}(t) = n_{\rm c}(t) \left|\int dx\, \varphi^{*}_{i}(x) \psi(x,t) \right|^2$ with $i=\pi$ and $0$, respectively. The qualitative behavior of the population dynamics in the experiment [Fig.~S6(a) of ref.~\cite{Lai2007nature}] is successfully reproduced by the collective transition process (solid curves). The population of the $\pi$-state initially increases with the pump pulse. At $t \approx 15\tau$, the density of the $\pi$-state reaches a critical value for dynamical instability, triggering its decay to the zero-state. Subsequently, the population of the zero-state surpasses that of the $\pi$-state, and becomes dominant.

In the experiment, the population of the zero-state starts to increase before that of the $\pi$-state reaches the maximum [see Fig.~S6(a) of ref.~\cite{Lai2007nature}]. By including the contribution of the single-particle transition [dashed curves in Fig.~\ref{fig:modetrans_pulse}(b)], the increase of the zero-state population at earlier time before the onset of the collective mode transition is reproduced. However, the change of $n_i$ by the single-particle transition is small, and it is thus concluded that the collective transition process plays a key role in the mode transition in the experiment~\cite{Lai2007nature}. Here, we remark that the mode-competition description based on rate equations of single-particle transitions presented in ref.~\cite{Lai2007nature} does not properly reproduce the mode-switching phenomenon observed in the experiment. We find that the value of the initially pumped exciton-polaritons employed in that calculation (denoted by $n_3$) is treated as a fitting parameter, and is inconsistent with the actual density of exciton-polaritons in the experiment.

Finally, we discuss the mode-switching phenomenon, i.e., the pump-power dependence of the integrated lower-polariton intensities of the $\pi$- and zero-states. Figure~\ref{fig:modetrans_pulse}(c) shows the cumulative density $\int n_{i}(t)\, dt$ (over $t=0$ to $50\tau$) of the $\pi$-state (red squares) and zero-state (blue circles) for various pumping strengths. For the $\pi$-state, its population is given by $n_\pi(t) = n_{\rm c}(t) |\int dx\, \varphi_{\pi}^*(x) \psi(x,t)|^2$, whereas for the zero-state population $n_0(t)$, the number of particles transferred from the $\pi$-state through the single-particle transition process is added to $n_{\rm c}(t) |\int dx\, \varphi_{0}^*(x) \psi(x,t)|^2$ (see Methods).

Here, we see that the accumulated density of the $\pi$-state is surpassed by that of the zero-state for sufficiently large pumping strength, and the mode-switching phenomenon observed in the experiment [Fig.~S5(a) of ref.~\cite{Lai2007nature}] is well reproduced. The interchange of the dominant state can be explained by the onset of the dynamical instability at large pumping strength $\overline{P}$ as follows. For small $\overline{P} \lesssim 3\ \text{ps}^{-1}$, the average density of the $\pi$-state is so low that the collective mode transition does not occur and only the single-particle transition contributes to the accumulated density of the zero-state. However, at larger $\overline{P} \gtrsim 3\ \text{ps}^{-1}$, where the $\pi$-state's density exceeds the threshold of the dynamical instability, the collective transition to the zero-state happens, which contributes to the accumulated density of the zero-state. For even larger $\overline{P}$, while the accumulated density of the $\pi$-state is saturated near the value for the critical case, that of the zero-state increases with the pumping strength and finally surpasses the former.

\bigskip
\noindent\textbf{Conclusion.}
In this work, we have revisited the mode-switching phenomenon observed in the experiment of ref.~\cite{Lai2007nature} and have clarified its mechanism. Unlike an accepted interpretation of this phenomenon based on single-particle transition processes~\cite{Lai2007nature, Deng2010RMP}, we have shown that the phase unwinding processes of the condensate wave function by topological defects play a key role and have found its evidence in the experimental data of the momentum distribution. Control of the state in open quantum systems is the research front line of quantum control for prospective quantum technology~\cite{Koch2022epjquantumtech}. Our work clarifying the mechanism of the mode switching phenomenon opens a route towards active control of the mode switching by manipulating the topological defects, which may be useful for future quantum polaritonic devices~\cite{Moxley2021emergentmat}. Furthermore, our work not only sheds light on the understanding of the mode-switching phenomenon in exciton-polariton condensates, but also inspires fresh insights into the collective dynamics of topological states in driven-dissipative quantum systems.

\bigskip
\noindent{\large\textbf{Methods}}\\
Here, we detail the methods and techniques used for the simulations discussed in the previous sections.


\bigskip
\noindent\textbf{Parameter estimate.}
Here we estimate the parameters in the gGPE (\ref{eq:ggpe}), based on the ones used in the experiment of ref.~\cite{Roumpos2011natphys}. Since this experiment was conducted by the same group of ref.~\cite{Lai2007nature} with a similar setup, their parameters are natural choice for our simulations in the present work.

In the theoretical discussion of ref.~\cite{Roumpos2011natphys}, also in Refs.~\cite{Ma2015PRB, Chestnov2016PRB}, the system is described by the coupled driven-dissipative GP model \cite{Wouters2007PRL, Carusotto2013RMP}. This model consists of the GP equation for the condensate wave function $\Psi$ of polaritons:
\begin{equation}
	\begin{aligned}
		i\hbar\frac{\partial\Psi(x,t)}{\partial t} =\Big[ &-\frac{\hbar^2}{2m} \nabla^2 + V_0^{\prime}\, \cos{\Big( \frac{2\pi}{d}x \Big)} + g_{\rm c}\, |\Psi(x,t)|^2 \\
		&+\frac{i\hbar}{2}(n_{\rm R}\, R- \gamma_{\rm c}) +n_{\rm R}\, g_{\rm R} \Big] \Psi(t)
	\end{aligned}
	\label{eq:partial_psi}
\end{equation}
coupled with the following rate equation describing the population dynamics of the reservoir excitons:
\begin{equation}
	\frac{\partial n_{\rm R}}{\partial t} = P' - n_{\rm R}\, \gamma_{\rm R} - n_{\rm R}\, R\, |\Psi(x,t)|^2.
	\label{eq:partial_n_R}
\end{equation}
Here, $n_{\rm R}$ is the density of the reservoir excitons; $P'$ is the gain rate of the reservoir excitons from the laser pump; $g_{\rm c}$ and $g_{\rm R}$ are the two-body interaction strength of the polaritons in the condensate and the one between the polariton and the reservoir exciton, respectively; $R$ is the scattering rate between the reservoir excitons and the polaritons in the condensate; $\gamma_{\rm c}$ and $\gamma_{\rm R}$ are the loss rates of the polaritons in the condensate and the reservoir excitons, respectively; and $V_0'$ is the depth of the external potential.
There~\cite{Roumpos2011natphys, Ma2015PRB, Chestnov2016PRB}, they took the following parameter values:
\begin{equation}
	\begin{aligned}
		g_{\rm c} &=6 \times 10^{-3}\ \text{meV}\, \mu \text{m}^2,\\
		g_{\rm R} &=2 g_{\rm c},\\
		\gamma_{\rm c} &=0.33\ \text{ps}^{-1},\\
		\gamma_{\rm R} &\approx 1.5 \gamma_{\rm c},\\
		R &=0.01\ \text{ps}^{-1}\, \mu \text{m}^2.
	\end{aligned}
	\label{eq:parameters}
\end{equation}
Our gGPE~(\ref{eq:ggpe}) is different from the above coupled GP model [i.e., Eqs.~(\ref{eq:partial_psi}) and (\ref{eq:partial_n_R})], but one can identify the following correspondence relations between the parameters in the two models from the homogeneous steady state.

In the coupled GP model [Eqs.~(\ref{eq:partial_psi}) and (\ref{eq:partial_n_R})], the density $n_{\rm R}$ of the reservoir excitons does not change in time in the steady state, and thus we get
\begin{equation}
	n_{\rm R} = \dfrac{P'}{\gamma_{\rm R} + R\, |\Psi|^2}
	\label{eq:n_R}
\end{equation}
from Eq.~(\ref{eq:partial_n_R}). When the system reaches the steady state, the imaginary terms on the right hand side (rhs) of Eq.~(\ref{eq:partial_psi}) should vanish for the homogeneous system:
\begin{equation}
	n_{\rm R} R - \gamma_{\rm c}=0.
	\label{eq:n_R2}
\end{equation}
From Eqs.~(\ref{eq:n_R}) and (\ref{eq:n_R2}), we obtain
\begin{equation}
	\dfrac{P'\, R}{\gamma_{\rm R} + R\, |\Psi|^2} = \gamma_{\rm c}.
\end{equation}
This equation can be rewritten as
\begin{equation}
	\begin{aligned}
		R\, |\Psi|^2 &= \dfrac{P'\, R}{\gamma_{\rm c}}-\gamma_{\rm R}\\
		&= \Big( \dfrac{P'\, R}{\gamma_{\rm c}} - \gamma_{\rm R} + \gamma_{\rm c} \Big) - \gamma_{\rm c}.
	\end{aligned}
	\label{eq:Rpsi}
\end{equation}
Similarly, in the gGPE~(\ref{eq:ggpe}), their imaginary terms also vanish in the homogeneous steady state. This reads the following relation:
\begin{equation}
	\eta\, |\Psi(x,t)|^2 = P - \gamma.
	\label{eq:etapsi}
\end{equation}

Since $\gamma_{\rm c}$ and $\gamma$ have the same physical meaning, i.e., the loss rate of the polaritons in the condensate, it is reasonable to regard that $\gamma_{\rm c}$ and $\gamma$ in the two models are identical. Thus, we have the following correspondence relation:
\begin{equation}
	\gamma_{\rm c} \ \longrightarrow \ \gamma.
	\label{eq:map_gamma}
\end{equation}
Comparing Eqs.~(\ref{eq:Rpsi}) and (\ref{eq:etapsi}) together with Eq.~(\ref{eq:map_gamma}), we can obtain
\begin{equation}
	R \longrightarrow \ \eta ,
	\label{eq:map_eta}
\end{equation}
\begin{equation}
	\dfrac{P'\, R}{\gamma_{\rm c}} - \gamma_{\rm R} + \gamma_{\rm c} \ \longrightarrow \ P.
	\label{eq:map_P}
\end{equation}
Finally, remaining correspondence relations between Eq.~(\ref{eq:partial_psi}) and Eq.~(\ref{eq:ggpe}) are
\begin{equation}
	g_{\rm c} \ \longrightarrow \ g ,
	\label{eq:map_g}
\end{equation}
\begin{equation}
	V_0' \ \longrightarrow \ V_0.
	\label{eq:map_V_0}
\end{equation}
Note that the term of $n_{\rm R} g_{\rm R}$ on the rhs of Eq.~(\ref{eq:partial_psi}) coming from the interaction between the polaritons in the condensate and the reservoir excitons is position-independent and just gives a shift in the energy reference point. Therefore, this term is irrelevant to the dynamics of the system, and there is no need to take it into account in identifying relations~(\ref{eq:map_g}) and (\ref{eq:map_V_0}). Taking the parameter values in Eq.~(\ref{eq:parameters}) for the coupled GP model, we can obtain their corresponding parameter values for the gGPE~(\ref{eq:ggpe}) via the relations~(\ref{eq:map_eta})--(\ref{eq:map_V_0}).

In addition, in the coupled GP model [Eqs.~(\ref{eq:partial_psi}) and (\ref{eq:partial_n_R})], the pumping threshold $P'_{\text{th}}$ to generate a polariton condensate in the homogeneous system given by~\cite{Wouters2007PRL, Ma2015PRB, Chestnov2016PRB}
\begin{align}
	P'_{\text{th}} = \gamma_{\rm R}\gamma_{\rm c}/R
	\label{eq:p'th}
\end{align}
is usually taken as a reference value for the pumping strength $P'$. On the other hand, in the gGPE (\ref{eq:ggpe}), the loss rate $\gamma$ is usually taken for the unit of $P$. Since the dimensions of the pumping strengths $P'$ and $P$ in the two models are different, it is convenient to have a relation between $P'/P'_{\text{th}}$ and $P/\gamma$. From Eq.~(\ref{eq:map_P}) with $\gamma_{\rm R} \approx 1.5 \gamma_{\rm c}$ in Eq.~(\ref{eq:parameters}), we obtain
\begin{equation}
	\dfrac{P'}{P'_{\text{th}}} = \dfrac{2}{3} \dfrac{P}{\gamma} +\frac{1}{3}.
	\label{eq:p'}
\end{equation}

\bigskip
\noindent\textbf{Bloch band structure of the steady state.}
To study the mode transition, first we calculate the steady Bloch states and obtain the $\pi$-state and zero-state. We set $\Psi(x,t) = \sqrt{n_{\rm c}}\, \varphi(x)\, e^{ikx-i\mu t}$, where $k$ is the quasimomentum, $\mu$ is the chemical potential, $\varphi(x)$ is the normalized condensate wavefunction of the steady state, $n_{\rm c} \equiv L^{-1}\int_{-L/2}^{L/2} |\Psi|^2 dx$ is the average density of the polariton condensate, and $L$ is the length of the system. Assuming that $ \varphi(x) $ can be expanded in the Fourier series:
\begin{equation}
	\begin{aligned}
		\varphi(x)=\sum\limits_{\nu =-{\infty}}^{+\infty} a_{\nu}\, e^{ i\frac{2\pi \nu}{d} x},
	\end{aligned}
\end{equation}
the gGPE~(\ref{eq:ggpe}) can be recast into a nonlinear eigenvalue problem for the coefficients $\{ a_{\nu} \}$, which is to be solved iteratively in a self-consistent manner. In the actual numerical calculations, the summation with respect to $\nu$ is taken up to a finite but sufficiently large cutoff value $\pm \nu_{\rm max}$ to get convergence. Note that the chemical potential must be real in the steady state \cite{Chestnov2016PRB}. Otherwise, the state will decay or grow in time because the pump $P$ is not balanced with the loss $\gamma$ and the gain saturation $\eta$. To obtain a real chemical potential $\mu$, we need to  adjust the average condensate density $n_{\rm c}$ to get the balance between the pump and loss.

\begin{figure}[tb!]
	\centering
	\includegraphics[width=1\linewidth]{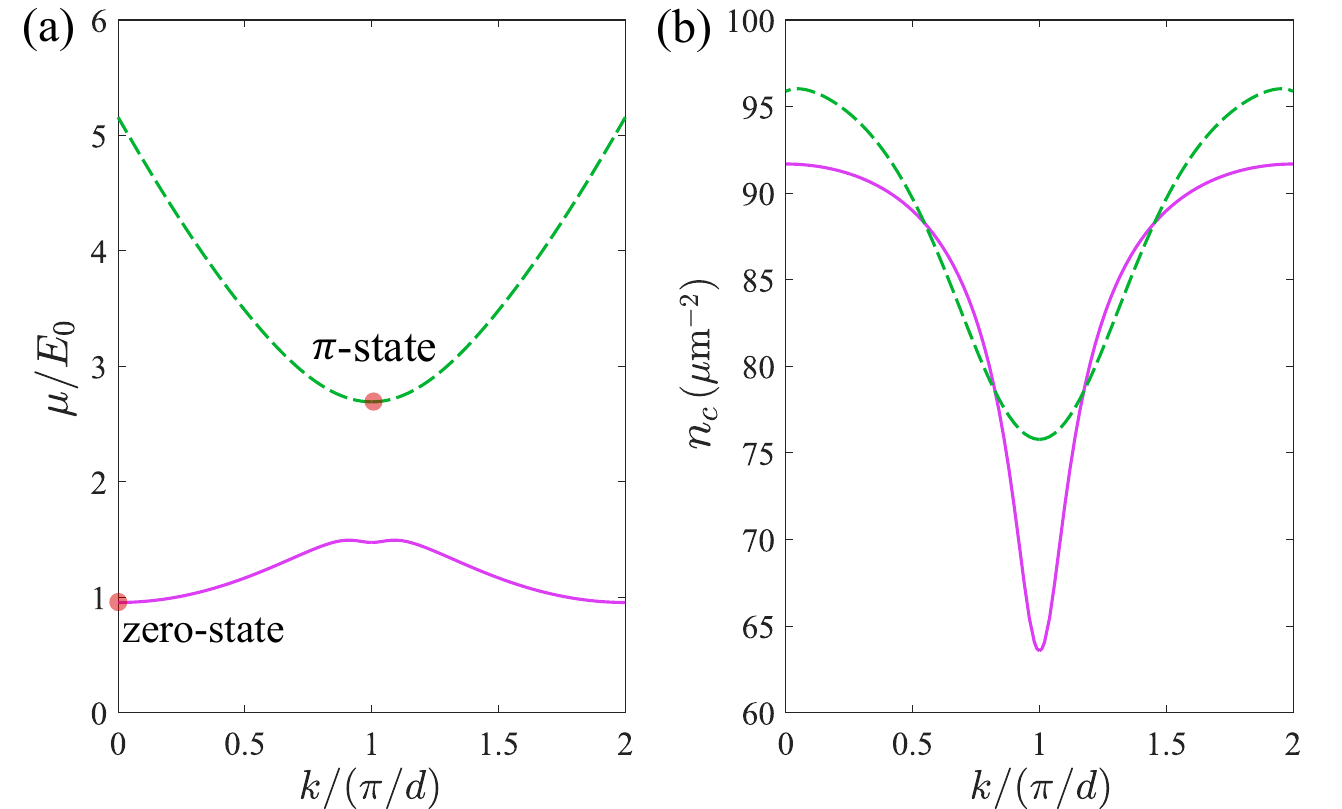}
	\caption{{\bf Bloch band structure and the $\pi$- and zero-states.}
	(a) Bloch band structure and (b) the corresponding average condensate density of the steady state. The first band and the second band are shown by the magenta solid curve and the green dashed curve, respectively. The red dots in (a) mark the $\pi$-state and the zero-state.}
	\label{fig:S1}
\end{figure}

The resulting Bloch band structure is shown in Fig.~\ref{fig:S1}(a). The magenta solid curve and the green dashed curve show the first and the second band, respectively, and the corresponding average condensate densities $n_{\rm c}$ are shown in Fig.~\ref{fig:S1}(b). The $\pi$-state refers to the state at the bottom of the second band with the quasimomentum $k=\pi/d$, and the zero-state refers to the state at the bottom of the first band with $k=0$.

The parameter values for our calculations of the gGPE are given by the correspondence relations~(\ref{eq:map_eta})--(\ref{eq:map_V_0}) with the parameter values (\ref{eq:parameters}). Namely, $g=6 \times 10^{-3} \text{meV}\ \mu \text{m}^2$, $\gamma=0.33\ \text{ps}^{-1}$, and $\eta=0.01\ \text{ps}^{-1} \mu \text{m}^2$. In addition, for the particular example of Fig.~\ref{fig:S1}, we take the potential depth $V_0/\hbar=1\ \text{ps}^{-1}$ and the pumping strength $P=4\gamma$, which corresponds to the pumping strength $P'=3P'_{\text{th}}$ in the coupled GP model [Eqs.~(\ref{eq:partial_psi}) and (\ref{eq:partial_n_R})] according to Eq.~(\ref{eq:p'}).

\bigskip
\noindent\textbf{Simulation of the population dynamics of the $\pi$- and zero-state under a pump pulse.}
The following two sections are about the details for generating Fig.~\ref{fig:modetrans_pulse}.

Using the gGPE~(\ref{eq:ggpe}), we simulate the experiment on the population dynamics of the $\pi$- and zero-state shown in Fig.~S6(a) of ref.~\cite{Lai2007nature}. In the simulation, we use the system with $10$ unit cells from $x=-5d$ to $5d$ under the periodic boundary condition, and the separation $\Delta x$ between the grid points is taken to be $\Delta x = 0.01d$.

Here, we take the potential depth $V_0/\hbar=0.1\ \text{ps}^{-1}$, which is comparable to the experimental value~\cite{Lai2007nature}, and the other parameters (except the pumping strength $P$) are the same as in the case of Fig.~\ref{fig:S1}. We first prepare a stationary $\pi$-state with a very small average condensate density as an initial seed by setting the pumping strength just above the threshold. Specifically, $P=1.15 \gamma$, which corresponds to $P'=1.1P'_{\text{th}}$ according to Eq.~(\ref{eq:p'}). Then, we perform the time evolution with a pump pulse starting from this initial $\pi$-state with a tiny delta noise added at $x=0$ (as a perturbation to the system). The temporal profile of the pump pulse $P(t)$ is given by
\begin{equation}
	P(t) =P(0) + \Delta P \times \left[ \mathrm{e}^{-(t-t_{\rm c})^2/t_{\rm w}^2} -  \mathrm{e}^{-t_{\rm c}^2/t_{\rm w}^2} \right]
	\label{eq:p_t}
\end{equation}
for $0 \le t \le 2t_{\rm c}$ and $P(t)=0$ for $t>2t_{\rm c}$ with the center of the pulse $ t_{\rm c}=18\tau $ ($ \tau \approx 1.24\ \text{ps}$) and the width of the pulse $ t_{\rm w}=10\tau $. The parameter $ \Delta P $ characterizing the height of the pulse is $ \Delta P = 15\gamma$ and $ P(0)=1.15\gamma$. According to Eq.~(\ref{eq:p'}), this $P(t)$ corresponds to
\begin{equation}
	P'(t)=1.1P'_{\text{th}}+10P'_{\text{th}}\times \left[ \mathrm{e}^{-(t-18\tau)^2/(10\tau)^2} -  \mathrm{e}^{-(18\tau)^2/(10\tau)^2} \right]
	\label{eq:p'_t}
\end{equation}
for $0 \le t \le 36\tau$ and $P'(t)=0$ for $t>36\tau$.

Let us remark on the duration and height of the pump pulse $P(t)$ used in the simulation. Note that the pumping strength $P(t)$ in the gGPE~(\ref{eq:ggpe}) is not equivalent to the power of the laser pulse in the experiment. Instead, $P(t)$ in the gGPE~(\ref{eq:ggpe}) is the gain rate from the exciton reservoir supplying particles to the condensate. Thus, the time scale of the population dynamics of the condensate is directly determined by the width and duration of $P(t)$. In the actual experiment, on the other hand, it takes some time for the polariton condensate to respond to the laser pulse. Namely, the laser pulse first creates high-energy excitons and the exciton reservoir will be replenished by these excitons after their relaxation. As a consequence, the time scale of the population dynamics of the condensate ($\sim 50~\text{ps}$) is much larger than the duration of the laser pulse ($2.5~\text{ps}$) as can be observed in Fig.~S6(a) of ref.~\cite{Lai2007nature}. Therefore, $P(t)$ in the gGPE can be very different from the temporal profile of the power of the pumping laser pulse in the experiment. In our simulation, we employ $P(t)$ given by Eq.~(\ref{eq:p_t}) with a longer duration ($36\tau \approx 44~\text{ps}$) and larger values for $t_{\rm c}$ and $t_{\rm w}$ to reproduce the population dynamics close to the one in the experiment.

Besides, it is noted that the pumping threshold $P'_{\text{th}} = \gamma_{_R} \gamma_{\rm c} / R$ given by Eq.~(\ref{eq:p'th}) is the one for the continuous and constant pump in the homogeneous system. The threshold for the pulse pump to generate a condensate in an inhomogeneous system is different from this value. To bridge the two cases, we introduce the average pumping strength $\overline{P'}$ defined by the time-average of $P'(t)$ over the duration $T$ of the pulse
\begin{align}
	\overline{P'} \equiv \frac{1}{T} \int_0^{T}dt\, P'(t)
\end{align}
as an effective value for the case of continuous constant pump. Then, the Gaussian pump pulse $P'(t)$ given by Eq.~(\ref{eq:p'_t}) with $T=36\tau$ corresponds to the effective pumping strength $\overline{P'}\approx 5.6P'_{\text{th}}$ for the continuous and constant pump. [Correspondingly, the average pumping strength $\overline{P} \equiv T^{-1} \int_0^T dt\, P(t)$ for $P(t)$ given by Eq.~(\ref{eq:p_t}) is about $7.9 \gamma$.] This effective pumping strength is comparable to the pumping power ($4$ times the threshold pumping power) used in the experiment~\cite{Lai2007nature}.

\bigskip
\noindent\textbf{Contribution of the single-particle transition from the $\pi$- to zero-state.}
To examine the contribution of the single-particle transition in the population dynamics and in the mode competition phenomenon, we incorporate the single-particle transition from the $\pi$- to zero-state by the stimulated process in our simulation.

Since our system initially starts to have a condensate in the $\pi$-state, the polaritons in the condensate are transferred to non-condensate polaritons in the zero-state by the stimulated transition. Therefore, the effect of this process can be incorporated by introducing an additional loss term in the gGPE describing the loss of the particles in the condensate in the $\pi$-state. The increase of the zero-state population by the single-particle transition can be evaluated by the number of particles lost from the $\pi$-state. The form of this additional loss term is set to reproduce the rate equation employed in ref.~\cite{Lai2007nature}. Namely, the loss rate of the $\pi$-state population by the stimulated transition from the $\pi$- to zero-state is proportional to $n_{\pi} (n_0 + 1)$, where $n_i$ with $i=\pi$ and $0$ is the population of the $\pi$-state and zero-state, respectively, given by $n_{i}(t)=n_{\rm c}(t) |\int dx \varphi_{i}^*(x) \psi(x,t)|^2 $. The reasonable form of the loss term is
\begin{equation}
	H_\text{loss}(t)=-\frac{i\hbar}{2}  \xi\, \gamma_{\pi \rightarrow 0}\, \left[  n_{0}(t)+1\right]\, n_{\pi}(t)/n_{c}(t),
	\label{eq:H_loss}
\end{equation}
where $\gamma_{\pi \rightarrow 0}$ is the transition rate from the $\pi$- to zero-state and $\xi$ is the correction factor for the loss rate introduced to rescale the difference in the average number density, which we shall discuss in detail later. With this additional loss term, the dynamics of the condensate is described by
\begin{equation}
	\begin{aligned}
		i\hbar \dfrac{\partial \Psi(x,t)}{\partial t}=  \left( H + H_\text{loss} \right)\Psi(x,t),
		\label{eq:ggpe2}
	\end{aligned}
\end{equation}
where $H$ is the generalized GP Hamiltonian of the original gGPE~(\ref{eq:ggpe}),
\begin{align}
	H=&-\frac{\hbar^2}{2m} \nabla^2 + V_0\cos{\Big( \frac{2\pi}{d}x \Big)} +g|\Psi(x,t)|^2 \nonumber\\
	&+\frac{i\hbar}{2}\left(P - \gamma - \eta|\Psi(x,t)|^2\right).
\end{align}
One can readily see that the modified gGPE~(\ref{eq:ggpe2}) with the additional loss term leads to the desired form of the rate equation of $n_{\rm c}$:
\begin{align}
	\frac{\partial }{\partial t}n_{\rm c}=&\frac{\partial }{\partial t} \dfrac{1}{L}\int_{-L/2}^{L/2}  |\Psi(x,t)|^2\, dx \nonumber\\
	=& \dfrac{1}{i\hbar}\dfrac{1}{L}\int_{-L/2}^{L/2}\left[ \Psi(x,t)^* (H+H_\text{loss}) \Psi(x,t) -\text{c.c.} \right]\, dx \nonumber\\
	=& \dfrac{1}{L}\int_{-L/2}^{L/2} \left(P - \gamma - \eta|\Psi(x,t)|^2 \right) |\Psi(x,t)|^2 dx \nonumber\\
	&- \xi\, \gamma_{\pi \rightarrow 0}\, \left[n_{0}(t)+1\right]\, n_{\pi}(t).
	\label{eq:nc}
\end{align}
The second term on the rhs describing the loss due to the stimulated transition from the $\pi$- to zero-state indeed has the same form as the one in the rate equation in ref.~\cite{Lai2007nature}. The first term on the rhs is due to the pump, loss, and gain-saturation of the system, which has already been taken into consideration in the simulation by the original gGPE~(\ref{eq:ggpe}). The population dynamics of the $\pi$-state by Eq.~(\ref{eq:ggpe2}) is shown by the magenta dashed line in Fig.~\ref{fig:modetrans_pulse}(b).

From Eq.~(\ref{eq:nc}), one can readily see that the increase in the zero-state population by the single-particle transition from the $\pi$-state is evaluated as follows:
\begin{align}
	\Delta n_{\pi \rightarrow 0}(t) = \int_0^t dt\, \xi\, \gamma_{\pi \rightarrow 0}\, \left[  n_{0}(t)+1\right]\,  n_{\pi}(t).
\end{align}
We add $\Delta n_{\pi \rightarrow 0}(t)$ to the density $n_0(t)$ of the zero-state, and meanwhile, the accumulated density $\Delta n_{\pi \rightarrow 0}(t)$ in the zero-state by the single-particle transition also decays with time at a rate $\gamma_{\rm c}=0.33 \text{ps}^{-1}$ as the polaritons in the condensate. The resulting population dynamics of the zero-state taking account of the contribution of $\Delta n_{\pi \rightarrow 0}(t)$ and its decay is shown by the cyan dashed line in Fig.~\ref{fig:modetrans_pulse}(b).

Finally, we discuss the correction factor $\xi$ for the loss rate $\gamma_{\pi \rightarrow 0}$. In ref.~\cite{Lai2007nature}, the values of the loss rates in the rate equation are determined to reproduce the experimental results (i.e., Figs.~S5 and S6 in \cite{Lai2007nature}) for an assumed value of the initial population of the reservoir excitons. Therefore, the appropriate value of $\gamma_{\pi \rightarrow 0}$ in front of the term for the stimulated transition changes by the actual population of the system because this term is nonlinear in the population. To use this coefficient $\gamma_{\pi \rightarrow 0}$ in our simulations with a different value of the population, the value of $\gamma_{\pi \rightarrow 0}$ should be rescaled accordingly. In ref.~\cite{Lai2007nature}, $\gamma_{\pi \rightarrow 0}=0.04\ \text{ps}^{-1} \mu \text{m}^{2}$ was obtained for an assumed average density $30\ \mu \text{m}^{-2}$ of the initial reservoir excitons. In our simulations, where the typical condensate density is $\sim100\ \mu \text{m}^{-2}$, $\gamma_{\pi \rightarrow 0}$ should be rescaled by a factor of $\xi=30/100$.

\bigskip
\noindent{\large\textbf{Data availability}}\\
All the simulation data in this paper can be reproduced using the described methodology. The relevant experimental data are available upon reasonable request.

\bigskip
\noindent{\large\textbf{Code availability}}\\
The codes used to generate the figures are available upon reasonable request.

\bigskip
\noindent{\large\textbf{Acknowledgments}}\\
G.W. was supported by the NSF of China (Grants No.~12375039, No.~11975199, and No.~11674283), by the Zhejiang Provincial Natural Science Foundation Key Project (Grant No.~LZ19A050001), and by the Zhejiang University 100 Plan. Y.M. was supported by the JSPS KAKENHI (Grant No.~JP19H05791 and No.~JP19H05603).
N.Y.K. acknowledges the support of Industry Canada, the Ontario Ministry of Research \& Innovation through Early Researcher Awards (RE09-068) and the Canada First Research Excellence Fund-Transformative Quantum Technologies (CFREF-TQT).
T.B. was supported by the National Natural Science Foundation of China (62071301); NYU-ECNU Institute of Physics at NYU Shanghai; the Joint Physics Research Institute Challenge Grant; the Science and Technology Commission of Shanghai Municipality (19XD1423000,22ZR1444600); the NYU Shanghai Boost Fund; the China Foreign Experts Program (G2021013002L); the NYU Shanghai Major-Grants Seed Fund; Tamkeen under the NYU Abu Dhabi Research Institute grant CG008; and the SMEC Scientific Research Innovation Project (2023ZKZD55).

\bigskip
\noindent{\large\textbf{Author contributions}}\\
G.L., T.B. and G.W. conceived the project. N.Y.K. provided information of the experiment and the experimental data. G.L. conducted theoretical and numerical studies with support from Y.M., N.Y.K., T.B. and G.W. All authors participated in the discussions. G.L., T.B. and G.W. wrote the manuscript with input from the other authors. G.W. led the project.

\bigskip
\noindent{\large\textbf{Competing interests}}\\
The authors declare no competing interests.

\bigskip
\noindent{\large\textbf{References}}
\bibliography{polariton_refs}

\end{document}